\documentclass[twocolumn,floatfix, nofootinbib,prd,preprintnumbers,showpacs,showkeys,superscriptaddress,longbibliography]{revtex4-2}
\usepackage{slashed}
\usepackage{amsmath}
\usepackage{booktabs}
\usepackage{array}

\usepackage{graphicx} % Required for inserting images

\usepackage{mathrsfs}
\usepackage{amssymb}
\usepackage{amsmath}
\date{\today}
\usepackage{mathtools}
\usepackage{comment}

\usepackage[mathscr]{euscript}
\usepackage{amsmath}
\usepackage{graphicx}
\usepackage{dcolumn}
\usepackage{bm}
\usepackage{epsfig}
\usepackage{amssymb,latexsym,mathrsfs}
\usepackage{graphicx}
\usepackage{color}
\usepackage{hyperref} 
\usepackage{float}
\usepackage{diagbox}

\usepackage{textgreek}

\usepackage{tikz}

\usepackage{amsmath}
\usepackage{physics}
\usepackage{csquotes} % \textquote{ } -- for quotes
\newcommand*\diff{\mathop{}\!\mathrm{d}}

\hypersetup{
    colorlinks=true,
    linkcolor=red,
    citecolor=blue,
} 

\usepackage{amsmath}
\usepackage{amssymb}
\usepackage{subfigure}
\usepackage{url}
\usepackage{xcolor}
\usepackage{color}
\usepackage{soul}
\definecolor{amaranth}{rgb}{0.9, 0.17, 0.31}
\definecolor{purple(munsell)}{rgb}{0.62, 0.0, 0.77}
\definecolor{americanrose}{rgb}{1.0, 0.01, 0.24}
\definecolor{palatinateblue}{rgb}{0.15, 0.23, 0.89}
\definecolor{royalblue(web)}{rgb}{0.25, 0.41, 0.88}
\definecolor{hanpurple}{rgb}{0.32, 0.09, 0.98}
\definecolor{beaublue}{rgb}{0.74, 0.83, 0.9}
\definecolor{carminered}{rgb}{1.0, 0.0, 0.22}
\definecolor{brightpink}{rgb}{1.0, 0.0, 0.5}
\definecolor{vividviolet}{rgb}{0.62, 0.0, 1.0}
\hypersetup{ linktoc=all,
    colorlinks, linkcolor={palatinateblue},
    citecolor={brightpink}, urlcolor={amaranth}}

\usepackage{comment}

\newcommand{\be}{\begin{equation}}
\newcommand{\ee}{\end{equation}}
\newcommand{\bs}{\begin{split}} 
\newcommand{\bea}{\begin{eqnarray}}
\newcommand{\eea}{\end{eqnarray}}

\newcommand{\bes}{\begin{subequations}}
\newcommand{\ees}{\end{subequations}}

\newcommand{\bo}{\raise-1mm\hbox{\Large$\Box$}}

\usepackage{cleveref}
\crefname{equation}{Equation}{Equations}
\Crefname{equation}{Equation}{Equations}

\usepackage{orcidlink}

\begin{document}

\preprint{FTPI-MINN-25-10}
\preprint{UMN-TH-4504/25}
% Particle creation from entanglement entropy
% Michael R. R. Good, Evgenii Ievlev, and Eric V. Linder
% M. Good, E. Ievlev, and E. Linder

\title{Particle creation from entanglement entropy} 
\author{Michael R. R. Good\,\orcidlink{0000-0002-0460-1941}}
%\orcidlink{0000-0002-0460-1941}
\email{michael.good@nu.edu.kz}
\affiliation{Physics Department \& Energetic Cosmos Laboratory, Nazarbayev University,\\
Astana 010000, Qazaqstan.}
\affiliation{Leung Center for Cosmology and Particle Astrophysics,
National Taiwan University,\\ Taipei 10617, Taiwan.}
\affiliation{Beyond Center for Fundamental Concepts in Science, Arizona State University,\\ Tempe AZ 85287, USA.}

\author{Evgenii Ievlev\,\orcidlink{0000-0002-5935-4706}}

\email{ievle001@umn.edu}
\affiliation{William I. Fine Theoretical Physics Institute, School of Physics and Astronomy,
University of Minnesota, \\
Minneapolis, MN 55455, USA.}

\author{Eric V.\ Linder\,\orcidlink{0000-0001-5536-9241}  }
\email{evlinder@lbl.gov} 
\affiliation{Berkeley Center for Cosmological Physics \& Berkeley Lab, University of California,\\ Berkeley CA 94720, USA.}

\begin{abstract} 
We investigate how entanglement entropy can drive particle creation, deriving explicit relations between entropy and the radiated particle spectrum, the total number of particles, and the total energy. Particle production is computed for scenarios that include accelerated motion, black hole evaporation, and beta decay, validating against known results while also extending them. We focus primarily on the low-entropy limit (analogous to non-relativistic motion), but also examine cases of significant particle production arising from harmonic cycles. The results establish an explicit operational link between information flow and matter creation, providing a concrete demonstration of `it from bit’.

\end{abstract} 

\date{\today} 

\maketitle

%\tableofcontents

\section{Introduction}

The origin of particles in quantum field theory is often traced to the dynamics of spacetime horizons, accelerating observers, or boundary conditions, each of which leads to the production of radiation. 
Well-known examples include Hawking radiation \cite{Hawking1974_Nature,Hawking:1974sw}, arising from horizons in black hole spacetimes; the Davies--Unruh effect \cite{Davies:1974th,unruh76}, where uniformly accelerated observers detect a thermal bath in Minkowski space; and the Davies--Fulling effect \cite{DeWitt:1975ys,Fulling:1972md,Davies:1976hi,Davies:1977yv}, describing particle creation by accelerating mirrors. 

Recent works on black hole analogs, acceleration, and the dynamical Casimir effect, further explore the robustness of both thermal and nonthermal particle production and its intimate link with entropy \cite{Trunin:2021fom,Xie:2023wvu,Paston:2024nma,Ageev:2023hxe,Belfiglio:2023sru,Ageev:2022hqc,Belfiglio:2025hzo,Lynch:2023zll,Osawa:2024fqb,Lynch:2024gft}. 
In all these cases, radiation is tied to information flow and entropy balance, hinting that quantum information may be as fundamental as geometry or dynamics in governing particle creation. Motivated by this perspective, we ask whether quantum information itself can serve as a primary driver of radiation. 

Despite decades of progress in quantum field theory and quantum information, a persistent gap remains in understanding how matter arises from information, a central tenet of Wheeler’s “it from bit” vision \cite{Wheeler1990}. 
While entanglement entropy has become a cornerstone in connecting geometry and information in gravitational systems, there exists no known computation that directly relates particle number to entanglement entropy. 
The absence of such a connection is striking. If information truly gives rise to matter, then the creation of particles from the dynamics of entanglement should not only be possible but inevitable. 

In this paper, we address this problem directly by proposing a framework that links time-dependent entanglement entropy to particle production in quantum field theory, offering an operational route to count particles from a bit-based analysis. 
Specifically, we examine whether time-dependent von Neumann entropy can act as a source of quanta. 
Rather than treating entropy as an outcome inferred from radiation, we ask whether manipulating the information content of a system directly induces particle creation. 
We further analyze the particle and energy spectra, as well as the total particle number and total energy, under various entropy profiles, identifying when the resulting distributions remain well-behaved in the IR and UV. 

Using the moving mirror model as a testbed, we formulate a framework where $S(t)$, the von Neumann entanglement \cite{Holzhey:1994we}, directly determines the number of emitted particles. 
In Section~\ref{sec:formalism}, we present the mathematical formalism and highlight IR/UV subtleties. 
We then illustrate cases without radiation in Section~\ref{sec:expan}, followed by nontrivial scenarios with particle production, including analogs of black hole evaporation and electron acceleration in beta decay, in Section~\ref{sec:nontrivial}. 
Section~\ref{sec:harmonic} investigates cyclic entropy variations that yield arbitrarily large particle production. 
We conclude in Section~\ref{sec:concl} with a summary of the connections uncovered between information flow, entanglement, and particle creation. 
Throughout, we set $\hbar = k_\textrm{B} = c = 1.$

%%%%%%%%%%%%%%%%%%%%%%%% 
\section{Method} \label{sec:formalism} 

Our analysis is confined to nonrelativistic motion, which is equivalent to the regime of low entanglement entropy. Since entropy scales with rapidity (or velocity), small entropy corresponds to slow motion and thus a suppressed rate of particle creation. Working in this limit extracts essential physics of particle production while sidestepping complications from ultra-relativistic motion.

%%%%%%%%%%%%%%%%%% 
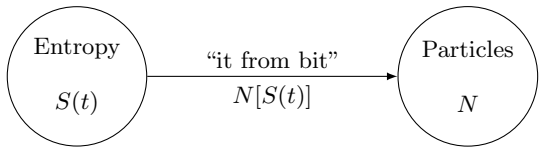
\begin{figure}
    \centering
    \begin{tikzpicture}[>=latex]
        % Nodes
        \node[draw,circle,align=center] (B) at (210:3) {Entropy\\\\$S(t)$};
        \node[draw,circle,align=center] (C) at (330:3) {Particles\\\\$N$};
        % Arrow with label
        \draw[->] (B) -- node[midway,above,sloped]{``it from bit"}
                 node[midway,below,sloped]{\small \(N[S(t)]\)} (C);
    \end{tikzpicture}
    \caption{A time-dependent entanglement entropy $S(t)$ may be used to determine $N$, the number of particles created.  }
    \label{fig:triality}
\end{figure} 

%%%%%%%%%%%%%%%%%%

%%%%%%%%%%%%%%%%%%%% 
\subsection{Entropy and Velocity}
The entanglement entropy of a moving mirror \cite{Holzhey:1994we} is\footnote{See also, e.g., section IV of \cite{Good:2020nmz}, appendix B of \cite{Fitkevich:2020okl}, section VII of \cite{Myrzakul:2021bgj}, below Eq. 14 of \cite{Bianchi:2014qua}, Eq. 8 of \cite{Chen:2017lum}, Eq. 14 of \cite{Good:2015nja}, C9 of \cite{mobmir}, etc.} 
\be \textrm{entropy} = -\frac{\textrm{rapidity}}{6} \approx -\frac{\textrm{velocity}}{6}\,, \label{entropy_velocity}\ee 
where the approximation sign denotes non-relativistic motion, within our scope of focus. 
The position of the mirror $z_t\equiv z(t)$ is then 
\be z_t = \int v_t \diff{t} = -6\int S_t \diff{t}.\ee
Therefore, the mathematical framework, at least in the moving mirror model, can be developed from two physically equivalent perspectives: (1) the mirror's equation of motion, $z_t$, or (2) the information encoded in the time-dependent von Neumann entanglement entropy, $S_t$.

%%%%%%%%%%%%%%%%%% 
\subsection{Particles and Particle-energy} 

Particle creation is described by 
the beta Bogolyubov coefficients of quantum creation/annihilation operators. In the moving mirror (accelerated boundary correspondence) picture, this takes a simple form for  nonrelativistic mirror motion; summing over both sides of the mirror yields  \cite{Mujtaba:2024vmf}:
\be |\beta_{pq}|^2 = \frac{4}{\pi}\ p q \,|z_\omega|^2, \label{betas}\ee
where $p$ and $q$ are the outgoing and ingoing radiation frequencies, $\omega = p+q$, and the Fourier transform is defined as
\begin{equation}
    z_\omega = \mathcal{F}_\omega z(t)= \frac{1}{\sqrt{2\pi}}\int_{-\infty}^\infty dt \, z(t) e^{-i\omega t}  \,.
\label{fourier_def_2}
\end{equation}
Using the property
\begin{equation}
    \mathcal{F}_\omega [ \dot{z}(t) ] = i \omega z_\omega,
\label{F_dot_z}
\end{equation}
we can write
\begin{equation}
    z_\omega = \frac{6 i}{\omega} S_\omega,
\label{z_S_omega}
\end{equation}
where $S_\omega$ is the Fourier transform of the entropy, and therefore 
\be |\beta_{pq}|^2 = \frac{144}{\pi}\ \frac{pq}{\omega^2}\ |S_\omega|^2\ . \label{betasw}\ee

To derive the formula for the particle count from the entropy 
we start from \cite{Mujtaba:2024vmf}
\begin{equation}
    N(p) = \int_0^\infty |\beta_{pq}|^2 \diff{q}\,,
\label{particlespectrum}
\end{equation}
and use Eq.~\eqref{betas} and Eq.~\eqref{z_S_omega} to obtain the particle spectrum
\begin{equation}
    N(p) = \frac{144}{\pi} \int \diff{q}  \, \frac{pq}{(p+q)^2} \left| S_{p+q} \right|^2\,,  
\label{particle_distr_from_S}
\end{equation} 
The total particle count is
\begin{equation}
    N  
        = \int_0^\infty N(p) \diff{p}
        = \frac{144}{\pi} \int \diff{p} \diff{q} \, \frac{pq \; \left| S_{p+q} \right|^2}{(p+q)^2}  \,. 
\label{totalparticles}
\end{equation}
We can actually perform one integration in Eq.~\eqref{totalparticles} explicitly.
Changing the integration variables to $\omega = p+q$ and $k=p-q$ with the Jacobian $1/2$, we obtain
\begin{equation}
\begin{aligned}
    N  
        &= \frac{144}{\pi} \int_0^\infty \diff{p} \int_0^\infty   \diff{q} \frac{pq}{ (p+q)^2 } |S_{p+q}|^2 \\
        &= \frac{18}{\pi} \int_0^\infty \diff \omega \int_{-\omega}^\omega \diff k  \frac{\omega^2 - k^2}{ \omega^2 } |S_{\omega}|^2 \,.
\end{aligned}
\end{equation}
Integrating over $k$ we arrive at
\begin{equation}
    N  
        = \frac{24}{\pi} \int_0^\infty \diff \omega \, \omega |S_{\omega}|^2 \,.
\label{totalparticles_compact}
\end{equation}

From the particle number distribution $N(p)$ we can also compute the total radiated energy,
\begin{equation}
\begin{aligned}
    E  
        &= \int_0^\infty p N(p) \diff{p} \\
        &= \frac{144}{\pi} \int_0^\infty \diff{p} \int_0^\infty   \diff{q} \, \frac{p^2 q}{(p+q)^2} \left| S_{p+q} \right|^2  \\
        &= \frac{12}{\pi} \int_0^\infty \diff \omega \, \omega^2 |S_{\omega}|^2 \, .
\end{aligned}
\label{particle_energy}
\end{equation}

Therefore, a time-dependent entanglement $S(t)$ may be used to determine $N$, the number of particles created. This is the central topic of investigation for this paper, with Figure~\ref{fig:triality} serving as the illustration.

We will use Eqs.~(\ref{particle_distr_from_S})--(\ref{particle_energy}) to compute the particle creation in a variety of situations. One particular case in Section~\ref{sec:bha} obtains the particle solution for a previously unsolved black hole analog, verifying that the total energy agrees with the result $E=M$ first solved in \cite{Good:2021ffo} using the usual stress-energy approach.

%%%%%%%%%%%%%%%%% 
\subsection{Power and Stress-energy} 

A complementary and more straightforward approach to calculate the energy bypasses the particle-spectrum calculation entirely: derivatives of $S(t)$ give the instantaneous power, and a final time integral yields the total radiated energy. While this method does not provide particle numbers or particle spectra, it offers a simple way to compute the energy of the radiation, which we will use throughout the paper to cross-check the energy carried by the particles.

Using (relativistic) Eq.~65 of \cite{Myrzakul:2021bgj}, the non-relativistic power emitted from both sides of the mirror is,
\be \text{Power} = \frac{6}{\pi} S'(\tau)^2 
\overset{\text{non-rel.}}{\approx}
\frac{6}{\pi} S'^{2}_t\,,\ee
while the relativistic expression Eq.~104 of \cite{Myrzakul:2021bgj} tells us the non-relativistic radiation reaction self-force magnitude,
\be \textrm{Force} = \frac{1}{\pi} S''(\tau) \overset{\text{non-rel.}}{\approx} \frac{1}{\pi}S''_t\,.\ee
The total stress-energy emitted may be found via time analysis only, using either the quantum power or the quantum reaction force; see, e.g., the classical analog via Eq.~(9.1.2) of \cite{Feynman:1996kb}:
\be E = \frac{6}{\pi}\int_{-\infty}^{+\infty} S'^{2}_t \diff{t} = -\frac{6}{\pi}\int_{-\infty}^{+\infty} S''_t S_t \diff{t}\,,\label{stress_energy} \ee 
with the total derivative vanishing. 
Under the right asymptotic behavior for $S(t)$, the particle-energy Eq.~\eqref{particle_energy} can be confirmed via the stress-energy, Eq.~\eqref{stress_energy}. 

By Parseval's theorem and the property Eq.~\eqref{F_dot_z}, one can immediately see that the total energy formulas Eq.~\eqref{particle_energy} and Eq.~\eqref{stress_energy} give the same result, consistent with the claim by Walker \cite{walker1985particle}.
At the same time, one can see that the total particle count $N$ from Eq.~\eqref{totalparticles_compact} cannot be written down in a simple way in terms of the time-dependent entropy $S(t)$, as such an expression would involve fractional derivatives.

%%%%%%%%%%%%%%%%%%% 
\subsection{Asymptotic and Causal Constraints} 

Regarding asymptotic behavior, there is a maximum entanglement entropy imposed by the physical necessity to use time-like worldlines, further constrained by our restriction to non-relativistic speeds, 
\be |\dot{z}(t)| \ll \textrm{speed of light}\quad \rightarrow \quad |S(t)| \ll 1/6\,.\ee
In addition, we will require that the particle count must not diverge. Thus, it is best to consider the trajectories that are asymptotic resting:
\be \textrm{Asymptotic rest:} \qquad \lim_{t\to\pm \infty} |\dot{z}(t)| = 0, \label{smooth}\ee
also expressed in terms of entanglement as (see Eq.~8 of \cite{Bianchi:2014vea}),
\be \textrm{Page curves \cite{Page:1993wv}:} \qquad \lim_{t\to\pm \infty} |S(t)| = 0, \label{smoothS}\ee
to ensure a finite particle count, see e.g., \cite{Good:2019tnf}.

The constraint $S_\textrm{max} \ll 1/6$ means we will be working with only a small amount of entanglement entropy. This is consistent with everyday life, where macroscopic systems exhibit negligible entanglement due to decoherence. Consequently, the particle creation rate remains small, reflecting minimal quantum excitation.

%%%%%%%%%%%%%%%%%% 
\subsection{IR convergence and purity}
\label{sec:IR_div_purity}

These bounds on \(S(t)\) inform the behavior of its Fourier transform \(S_\omega\), which we explore physically in this and the following subsection. 

First, the IR behavior of the entropy, Eq.~(\ref{z_S_omega}), at small frequencies is of interest. Suppose that the entropy in this limit obeys a power law
\begin{equation}
    S_\omega \sim \omega^{\gamma} \,, \quad
    \omega \to 0,
\end{equation}
with some exponent $\gamma$. 
Then, the IR behavior of the particle count integral in Eq.~\eqref{totalparticles_compact} is given by
\begin{equation}
    N \sim \int  \diff{\omega} \, \omega^{1 + 2 \gamma}.
\end{equation}
One can see that it is IR convergent only provided that
\begin{equation}
    \gamma > -1\,.
\label{particle_convergence_condition}
\end{equation}
Similarly, the total energy Eq.~\eqref{particle_energy} is IR convergent if $\gamma > -3/2$.
Note that if Eq.~\eqref{particle_convergence_condition} holds, the energy spectrum $E(\omega)$ itself is finite at $\omega \to 0$.

The question of IR convergence is an important one, and occurrences of IR divergencies signify some deep physical phenomenon.
On the contrary, UV divergencies often (but, admittedly, not always) signify some difficulties in the microscopic description that can be ignored in the long-distance effective theory.

To illustrate the importance of the IR physics, let us imagine that in our system, condition Eq.~\eqref{particle_convergence_condition} is violated, so that the energy spectrum $E(\omega)$ diverges at small frequencies, and the total number of radiated particles is infinite.
For example, suppose that $\gamma = -1$.
This suggests that our system is not entirely pure, i.e., some entropy remains unbalanced, and the system fails to return to its initial value.
Indeed, if the entropy does not return to its initial value,
\begin{equation}
    S(t \to +\infty) - S(t \to -\infty) = \Delta S \neq 0,
\label{purity_violation}
\end{equation}
then one can show (see Appendix~\ref{sec:math}) that in the IR the Fourier transform of the entropy satisfies
\begin{equation}
    \abs{ S_\omega } \approx \frac{ 2 \abs{\Delta S} }{ \sqrt{2\pi} } \frac{1}{\omega} \quad
    \text{at } \omega \to 0^+\,, 
\label{S_omega_IR_div}
\end{equation}
thus indeed yielding the entropy exponent $\gamma = -1$.

Note that, since the Fourier transform of a power law is a power law, the constraint Eq.~\eqref{particle_convergence_condition} on $S_\omega$ is saying that $S(t) \sim \text{const} + t^{\alpha=-(\gamma+1)<0}$, closely related to the requirement of Eq.~\eqref{smooth}. 
The constant term does not spoil purity as long as it is the same at the distant past and future, i.e., $\Delta S = 0$, since the Fourier transform of a constant is a delta function supported only at the zero frequency.
Thus, it slightly relaxes Eq.~\eqref{smooth}, and can be interpreted as a constant background contribution\footnote{See Eq.~(C8) in~\cite{mobmir} for the role of reference frame dependence. 
Preserving M\"obius symmetry requires using a frame-independent entropy definition, 
which ensures the particle count remains invariant under the symmetry transformation.}
 to the entropy, possibly originating from another subsystem.  Although an assumed power-law scaling in the IR is used for analytic clarity, other functional forms (e.g., exponential, logarithmic, or modulated behavior) could alter the convergence condition.  Since a power law scaling is a good ansatz, other such scalings are left for future analysis.

Therefore, using a power-law, we see that an IR divergence in the total energy signals that the system under consideration is not pure.

%%%%%%%%%%%%%%%%%% 
\subsection{UV divergence and discontinuity}
\label{sec:UV_disc}

For the UV behavior of the entropy, we can repeat the argument above for $\omega \to \infty$,  finding that the total particle count is convergent in the UV if the Fourier entropy falls off faster than $1/\omega$ at large $\omega$.
The borderline case, when the total particle count is logarithmically UV divergent, corresponds to $S_\omega \sim 1/\omega$ at large $\omega$.

The physics of such behavior can be easily understood based on the properties of the Fourier transform.
As an example, recall that if $S(t)$ is piecewise smooth but has finite jumps in isolated points, its Fourier transform falls off precisely as $1/\omega$ at large $\omega$ (the proof is analogous to that in Appendix~\ref{sec:math}).
Thus, a UV divergence of the total particle count indicates that, in the model of interest, entropy undergoes a discontinuity -- a finite jump at some moment in time.
Physically, this must be cured by some short-distance physics that makes the entropy smooth, but possibly rapid.
If $\Lambda$ is the UV scale of this microscopic physics, then the total particle count goes as $N \sim \ln\Lambda$.

To reiterate, if we encounter a system where the total particle count is divergent, this signifies a specific physical phenomenon.
The nature of these phenomena varies significantly depending on whether the divergence is IR or UV in nature.
Namely, an IR divergence indicates non-purity of the system (which cannot be cured by microscopic physics). In contrast, a UV divergence indicates a finite jump of entropy at some point in time (the microscopic physics can rectify this, giving a log dependence on the cutoff scale).

%%%%%%%%%%%%%%%%%%%% 
\section{Null Tests} \label{sec:expan} 

The concepts and relations in the previous section may suggest a novel role for entanglement in the production of particles. While in the moving mirror model $S(t)$ is tied to the mirror’s speed, in other systems (e.g.\  black holes or cosmological expansion), entanglement entropy might arise from different mechanisms. We pursue this further in the next section, but first want to establish that the entropy approach does not add spurious particle production. Therefore, we now examine three cases in which we expect no particle production to occur in the usual motion formalism, and confirm that the same holds in the entropy approach.

%%%%%%%%%%%%%%%%%%%%% 
\subsection{Static Mirror, Zero Entropy}

A mirror at fixed position \( z(t) = z_0 \) yields zero velocity and acceleration, so the entropy vanishes:
\begin{equation}
S(t) = 0\,.
\end{equation}
Hence, the time-integrated entropy and its Fourier transform are also zero, and by Eq.~(\ref{betasw}) the particle number does as well, 
\begin{equation}
N = 0\,.
\end{equation}
A static mirror radiates nothing, and so does a situation with zero entanglement entropy.

%%%%%%%%%%%%%%%%%%%%
\subsection{Constant Velocity Mirror, Constant Entropy}

Consider a low-speed boost of the previous mirror to one moving with constant velocity \( v_0 \). From Eq.~(\ref{entropy_velocity}), the entanglement entropy is proportional to the velocity,
\begin{equation}
S = -\frac{v_0}{6},
\end{equation}
and so is constant in time. 
Considering constant entanglement entropy situations, the Fourier transform $S_\omega$ is proportional to the Dirac delta function $\delta(\omega)$.
Therefore, it is thus supported only at zero frequency, which will make the particle Eq.~\eqref{totalparticles} and energy Eq.~\eqref{particle_energy} integrals vanish.

While physically consistent with Lorentz invariance, there is a mathematical subtlety: the integrals in Eqs.~\eqref{totalparticles} and \eqref{particle_energy} involve the square of \(S_\omega\), and if \(S_\omega \sim \delta(\omega)\), this square is ill-defined. This issue can be addressed using a regularization scheme, though selecting an appropriate one is not always straightforward, as some results may depend on the specific choice.

We found that the IR behavior of the total radiated energy and particles may depend on the details of how $S(t)$ is regularized for large $t$.
However, the most essential feature, namely the contribution from non-zero frequencies, turns out to be independent of regularization.
More precisely, our regularization prescription is as follows:
\begin{enumerate}
    \item Introduce a regularization 
    \item Compute particle distribution $N(p)$
    \item Remove regularization, discarding contributions that are supported at $p=0$ only
\end{enumerate}

For example, let us consider two choices of regularization.
One is the standard exponential cutoff:
\begin{equation}
    S^\text{reg}(t) = S(t) \, e^{ - \varepsilon |t|} ,
\label{S_reg_1}
\end{equation}
where $S(t)$ is the original entropy.
The other is an energy-dependent cutoff, meaning that we compute the Fourier transform as
\begin{equation}
    S^\text{reg}_\omega =  \frac{1}{\sqrt{2\pi}}\int_{-\infty}^\infty dt \, S(t) e^{ - \varepsilon | \omega | |t|}  e^{-i\omega t}.
\label{S_reg_2}
\end{equation}
Here, $\varepsilon$ is the regularization parameter; it will be sent to zero at the end.
The second regulator produces somewhat simpler formulas, but in principle, they are equally effective, and the statements below will be valid for both.

Now, let us treat the constant-entropy case.
We have
\begin{equation}
    S(t) = S^{(0)} = \text{const}.
\end{equation}
Performing a regularized computation of the particle distribution, we find (for either regularization)
\begin{equation}
    N^\text{reg}(p) = O(\varepsilon^2) \xrightarrow[\varepsilon \to 0]{} 0.
\end{equation}
Thus, the particle distribution vanishes after removing the regularization.

Constant entanglement, as with constant velocity motion, produces no energy and no particles, in agreement with the total stress-energy, Eq.~(\ref{stress_energy}).

\subsection{Uniformly Accelerated Mirror}
\subsubsection{Eternal}

In the relativistic case, it is well known that a mirror undergoing uniform proper acceleration does not radiate energy \cite{Birrell:1982ix}.
What does our non-relativistic particle creation approximation predict in this case?

Strictly speaking, the non-relativistic approximation is not applicable here, since for the constant-acceleration trajectory the velocity $\dot z = \kappa t$ becomes unbounded and inevitably relativistic.
Nevertheless, we can formally proceed using the same regularization scheme as in the previous subsection.

The trajectory we consider is
\begin{equation}
    z(t) = \frac{\kappa t^2}{2} \,, \quad
    t \in (-\infty,\infty),
    \label{smoothUA}
\end{equation}
where $\kappa$ is the constant acceleration.
The corresponding entropy is
\begin{equation}
    S(t) = S^{(0)} t \,, \quad S^{(0)} = \mathrm{const}.
    \label{S_uniform}
\end{equation}

The stress-energy Eq.~(\ref{stress_energy}) shows that the total energy vanishes (for a smooth, integrable function).  
Introducing a regularization effectively imposes a cutoff on the maximal velocity.  
For example, using Eq.~\eqref{S_reg_1}, the regularized motion has maximal speed scaling as
\begin{equation}
    \dot{z}^\mathrm{reg}_\mathrm{max} = O(1/\varepsilon) .
\end{equation}
Thus, the motion becomes truly non-relativistic only when $\varepsilon$ is sufficiently large.
Formally, however, we can take the limit $\varepsilon \to 0$ and find
\begin{equation}
    N^\mathrm{reg}(p) = O(\varepsilon^2) 
    \xrightarrow[\varepsilon\to 0]{} 0 .
\end{equation}
Both the particle number and total energy vanish, as expected for uniform eternal acceleration.

\subsubsection{Semi-eternal}

The situation changes dramatically if the uniform acceleration begins at a finite time.
Consider a mirror that is at rest for $t<0$ and accelerates for $t>0$:
\begin{equation}
    z(t) = 
    \begin{cases}
        0 \,, & t \leqslant 0 , \\[6pt]
        \dfrac{\kappa t^2}{2} \,, & t > 0 .
    \end{cases}
\end{equation}
The corresponding entropy is
\begin{equation}
    S(t) =
    \begin{cases}
        0 \,, & t \leqslant 0 , \\[6pt]
        S^{(0)} t \,, & t > 0 .
    \end{cases}
    \label{S_uniform_sing-br}
\end{equation}

Using the same regularization prescription, the particle spectrum no longer vanishes:
\begin{equation}
    N^\mathrm{reg}(p) = 
    \frac{12\sqrt{2}}{\pi^{3/2} p} (S^{(0)})^2 + O\!\left(\varepsilon^2\right),
    \qquad \varepsilon \to 0 .
\end{equation}
Even after removing the regulator ($\varepsilon=0$), $N(p)$ diverges as $p\to 0$, leading to a logarithmic infrared divergence.
This signals impurity, as discussed in Sec.~\ref{sec:IR_div_purity}.

For this reason, and the remainder of this paper, unless otherwise stated, we restrict attention to smooth, single-function trajectories.

%%%%%%%%%%%%%%%%%%%%%%%%%% 
\section{Entropic Particle Production} \label{sec:nontrivial} 

We here consider several cases illustrating how particle production can result from time-varying entanglement entropy, in very different scenarios, analogous to accelerated motion, black hole evaporation, and beta decay.

%%%%%%%%%%%%%%%%%% 
\subsection{Lorentzian Form}
\label{sec:lorentzian_mirror} 

As a demonstration of the connection between the energy and particles emitted to the maximum entanglement entropy, consider the Lorentzian function,
\be S(t) = -\frac{16 S_{\textrm{max}}}{3 \sqrt{3}} \frac{\kappa t}{ \left(\kappa^2 t^2+1\right)^2}\,,\ee
imposing $S_{\textrm{max}}\ll 1/6$; note $S(t) \to 0$ as $t\to\pm\infty$. 
The energy may be confirmed via Eq.~(\ref{particle_energy}) or Eq.~(\ref{stress_energy}), and they indeed agree with other,  
\be E = \frac{32}{3}\,\kappa\, S_{\textrm{max}}^2\,,\ee
giving us confidence that the particle count from Eq.~(\ref{totalparticles}), $N = 32 S_{\textrm{max}}^2/3$, is correct. This example\footnote{See a similar example in Appendix \ref{sec:slow_smooth_mirror}.} reveals a new relationship: within the small-entropy regime, the particle count is governed by the square of the maximum entanglement entropy.

\subsection{Black Hole Analog}
\label{sec:bha}
Having seen the connection between particle count and maximum entanglement above, we can extend this relationship in analogy by linking the particles and energy emitted to the mass of a black hole (via the entanglement entropy). 

Consider the black hole analog model of  \cite{Good:2021ffo} with its time-dependent entanglement entropy\footnote{Eq. 5 of \cite{Good:2021ffo} with $j=2$, $t_* = 96 \sqrt{2 \pi } M^3$, $\alpha_0 = 1/(4M)$.},
\be S_t = \frac{1}{6} \sinh ^{-1}\left(12 \sqrt{2 \pi } M^2 \Gamma \left[\frac{1}{2},\frac{t^2}{[96\sqrt{2\pi}\,M^{3}]^2  }\right]\right).\label{superS}\ee
This can be simplified by assuming $S_{\textrm{max}} \ll 1/6$, which allows us to drop the arc-hyperbolic sin,
\be S(t) \approx 2 \sqrt{2 \pi } M^2 \,\Gamma \left[\frac{1}{2},\frac{t^2}{[96\sqrt{2\pi}\,M^{3}]^2  }\right].\label{S_t_bh_analog_small}\ee
Eq.~(\ref{S_t_bh_analog_small}) is a good starting point for a time-dependent $S(t)$.  With this in hand, we can attempt to apply the entropic particle-energy creation formula, Eq.~(\ref{particle_energy}), and solve for the total energy $E$ of evaporation, verifying it against the stress-energy, Eq.~(\ref{stress_energy}).

We will need the tractable analytic Fourier transform:
\be S_\omega = \frac{8 M^2}{\omega} F[48 M^3 \sqrt{2 \pi } \omega],\ee
where the Dawson integral is $F[z] = e^{-z^2}\int_0^z e^{t^2}\diff{t}$. 
Making a change of variables for integration such that $q = \omega -p$, we rewrite the particle-energy integral Eq.~(\ref{particle_energy}) as:
$$ 
E = \frac{144}{\pi} \int_0^\infty \diff{\omega} \int_0^\omega \diff{p} \frac{p^2 (\omega - p)}{\omega^4} 64 M^4 F^2[48 M^3 \sqrt{2\pi} \omega]  .
$$ 
This can be analytically integrated and the result is 
\be E = M.\ee
This confirms the validity of Eq.~(\ref{particle_energy}) in this case, and demonstrates that starting with the entanglement entropy, the total energy of the particles from complete evaporation is just the mass of the black hole.  

Of course, one may also confirm $E=M$ via the much easier route of stress-energy, as given in Eq.~(\ref{stress_energy}). Still, the point of solving $E=M$ via particle-energy as given by Eq.~(\ref{particle_energy}) was to build confidence that we can use Eq.~(\ref{totalparticles}) to solve for the particle creation $N$. 

The particle count $N$, from Eq.~(\ref{totalparticles}),
$$
N = 
\frac{144}{\pi} \int_0^\infty  \int_0^\omega \frac{p^1 (\omega - p)}{\omega^4} 64 M^4 F^2[48 M^3 \sqrt{2\pi} \omega] \diff{p} \diff{\omega},
$$
admits an analytic evaluation. The result is
\be
N = \frac{768 C}{\pi} M^4 \label{catalan_particle_mass}
,\ee
where \( C \) is the dimensionless Catalan constant
\be
C = \sum_{n=0}^\infty \frac{(-1)^n}{(2n+1)^2} 
= \frac{1}{1^2} - \frac{1}{3^2} + \frac{1}{5^2} - \frac{1}{7^2} + \frac{1}{9^2} - \cdots , \ee
which gives $C \approx 0.915965$. Eq.~(\ref{catalan_particle_mass}) is a key advance in that it enables directly linking particle count to the mass of the black hole. In the limit \( M \to 0 \), particle production vanishes, but grows rapidly as \( M^4 \) for small mass (and hence small entropy).

In the present $1{+}1$ analog model, $M$ is not a physical mass but the intrinsic scale parameter of the entropy profile. Thus, the result $  N \sim M^{4}$ should be understood as scaling with this single model scale in natural units. 

By contrast, in the full $3{+}1$ black hole case Newton’s constant $G$ provides the extra dimensionful input that renders $N$ strictly dimensionless, allowing the result, Eq.~(\ref{catalan_particle_mass}), to be written in Planck units, e.g.\ $M/m_{\rm P}$ or $\kappa_{\rm P}/\kappa$,  
\be 
N = \frac{3C}{\pi}\,\frac{\kappa_{\textrm{P}}^{4}}{\kappa^{4}}, 
\qquad 
\kappa_{\textrm{P}}\equiv\sqrt{\frac{c^{7}}{\hbar G}} .
\ee
In our reduced flat-spacetime model, this gravitational input is absent, and the scaling can be expressed instead,
\be
N
= \frac{2C}{3\pi^{2}}(\alpha_{0}t_\star)^{2},
\ee
through the intrinsic time parameter $t_\star=96\sqrt{2\pi}\,M^{3}$. In this case, $\alpha_0 = 1/(4M)$, carries units of inverse time, while $t_\star$ is a time scale, so their product is dimensionless.

\subsection{Beta Decay Photons}\label{sec:beta_decay}

As a third demonstration, we investigate the actual motion of a particle, specifically an accelerating electron from beta decay. 
We will show that the relevant formulae consistently describe the photon radiation of beta decay in the non-relativistic regime. The total emitted photon energy radiated by the accelerated electron during beta decay with acceleration parameter $\kappa$, and asymptotic final speed $s$, is given by, e.g.\  \cite{Ievlev:2023inj},  
\be
E = \frac{e^2 \kappa}{24\pi} \left( \frac{\tanh^{-1}s}{s} - 1 \right) \approx \frac{e^2 \kappa}{24\pi} \left(\frac{s^2}{3}\right),\label{beta_decay_energy}
\ee
which agrees with known experimental results, see e.g., \cite{PhysRev.76.365,Lynch:2022rqy}.

Consider a classical point particle electron moving along the trajectory $z(t)$, \cite{Good:2022eub}, and generating a corresponding entanglement entropy,
\be
z(t) = \frac{s}{\kappa} \, W\left( e^{\kappa t} \right), \quad S(t) = \frac{s}{6} \left[ \frac{1}{W(e^{\kappa t}) + 1} - 1 \right],\label{betadecay_traj}
\ee
where \( W \) denotes the product logarithm, \( s \) is the final speed of the electron with \( s \ll 1 \), and \( \kappa \) is the acceleration scale. From this classical trajectory, it is straightforward to verify that the total radiated energy computed via the Larmor formula for power, $P = e^2 \ddot{z}^2/(6\pi)$, yields:
\[
E = \frac{e^2}{6\pi} \int_{-\infty}^{+\infty} \ddot{z}^2(t) \, \diff{t} = \frac{e^2 \kappa}{72\pi} s^2.
\]
Alternatively, by analogy with Eq.~\eqref{stress_energy}, the same result can be obtained from an entropic perspective,
\[
E = e^2 \frac{6}{\pi} \int_{-\infty}^{+\infty} \left[ S'(t) \right]^2 \diff{t} = \frac{e^2 \kappa}{72\pi} s^2.
\]
The next question we ask about the entropic particle creation process is whether the photons (emitted by this classically accelerated electron) have the same energy spectrum as in the usual analysis?

Using semi-classical physics, recall the non-relativistic particle spectrum (e.g., the classical analog of Eq.~(59) of \cite{Mujtaba:2024vmf} and Appendix~\ref{sec:expansion_beta_check} here), each quanta has energy $\omega$:
\be
N(\omega) = \frac{I(\omega)}{\omega}, \quad \text{where} \quad I(\omega) = \frac{e^2}{3\pi} \left| \ddot z(\omega) \right|^2. \label{Iw}
\ee 
Using the mathematical identity, Eq.~(32) of \cite{Mujtaba:2024vmf}, 
\[
\left| \mathcal{F}_\omega \left[ \partial_t^2 \frac{1}{\kappa} W(e^{\kappa t}) \right] \right|^2 = \frac{\omega/\kappa}{e^{2\pi \omega/\kappa} - 1},\ 
\]
and plugging into Eq.~(\ref{Iw}), we find
\[
I(\omega) = \frac{e^2 s^2}{3\pi} \frac{\omega/\kappa}{e^{2\pi \omega/\kappa} - 1}\ .
\]
Consequently, the total particle-energy is found to be
\[
E = \int_{0}^{\infty} I(\omega) \, \diff \omega = \frac{e^2 \kappa}{72\pi} s^2,
\]
in agreement with the known result, Eq.~(\ref{beta_decay_energy}).

Viewed from an entropic perspective, one considers $|\ddot z(\omega)|^2 = \omega^2 |\dot z(\omega)|^2 = 36\omega^2 |S_\omega|^2$; 
the particle spectrum is expressed as
\[
N(\omega) = \frac{12 e^2}{\pi } \, \omega \left| S(\omega) \right|^2.
\]
Using the entanglement entropy of Eq.~(\ref{betadecay_traj}), then 
\be
|S(\omega)|^2 = \frac{s^2}{36\omega\kappa} \cdot \frac{1}{e^{2\pi \omega /\kappa} - 1}\ , 
\ee
and indeed, we find that the photons are distributed in a 1-dimensional Planck spectrum: 
\be
N(\omega) = \frac{e^2 s^2}{3\pi \kappa} \cdot \frac{1}{e^{2\pi \omega /\kappa} - 1}\  \label{thermal_acceleration_particle_spectrum}.
\ee
The energy carried by the particles is
$$
E =  \int_{0}^{\infty}  \omega N_\omega \, \diff{\omega} = \frac{12 e^2}{\pi }\int_{0}^{\infty}  \,  \left| \omega S_\omega\right|^2 \, \diff{\omega} = \frac{ e^2 \kappa}{72\pi} s^2\,. \label{particle-energy-decay}
$$
This agrees with the total energy from Eq.~(\ref{beta_decay_energy}).  Although limited to the non-relativistic regime, this shows that the entropic particle perspective not only confirms the total particle-energy of beta decay but also reveals its thermality, Eq.~(\ref{thermal_acceleration_particle_spectrum}). 

As a side remark, note that the trajectory equation of motion Eq.~(\ref{betadecay_traj}) has a future-asymptotically constant velocity at speed $s$ (and hence asymptotic constant entropy). As such, there is an IR-divergence leading to infinite soft particles for the total count, $N$, when integrating over Eq.~(\ref{thermal_acceleration_particle_spectrum}),
\be N = \int_0^{\infty}N_\omega \diff{\omega} \sim \int_0^{\infty} \frac{\diff \omega}{e^{2\pi \omega /\kappa} - 1} \to  \textrm{divergent} \label{soft}.\ee
The entropic procedure used to determine energy Eq.~(\ref{beta_decay_energy}) is robust to this trait, extending its applicability beyond the black hole analog and Lorentzian trajectory discussed above. Those trajectories, being asymptotically static, produced finite particle counts.

%%%%%%%%%%%%%%%%%%%%%%%%%%%% 
\section{Harmonic Entropy: Large Particle Production} 
\label{sec:harmonic}

We have examined the cases of zero particle production, as well as the small but finite particle production from the mirror and black hole, and the small non-relativistic particle-energy production from the accelerating electron. We now consider the possibility of large particle production from a mirror's periodic motion.  

A moving mirror (boundary) radiates a certain amount of energy per one oscillation, so one that is in perpetual periodic motion therefore makes the total radiated energy and particles infinite.
In order to regularize the problem of computing particles and energy from the time-varying entropy, below we consider two possible scenarios: an oscillator that just stops after a certain number of periods (hard cutoff), and a damped oscillator (exponential cutoff).

%%%%%%%%%%%%%%%%%%%% 
\subsection{Finite number of oscillations}
\label{sec:harmonic-fin} 

Consider a continuous time-dependent entropy profile
\begin{equation}
S(t) = \begin{cases} 
    s \sin(\kappa t) , \quad &t \in [-\pi n /\kappa, \pi n /\kappa], \\
    0 \,, \quad &t \notin [-\pi n /\kappa, \pi n /\kappa],
    \end{cases}
\label{S_t_harmonic_sin}
\end{equation}
representing $n$ full wavelengths of harmonic entanglement. 
Such entropy corresponds to  harmonic motion that goes on for $n$ oscillations and then stops, with 
\begin{equation}
z(t) = \begin{cases} 
    - 6 (s/\kappa)\, \cos(\kappa t) , \quad &t \in [-\pi n /\kappa, \pi n /\kappa], \\
    (-1)^{n+1} \,6s/\kappa \,, \quad &t \notin [-\pi n /\kappa, \pi n /\kappa],
    \end{cases}
\label{z_t_harmonic_sin}
\end{equation}
following  Eq.~(\ref{entropy_velocity}).
Note that both $S(t)$ and $z(t)$ are both continuous. 

The Fourier transform of the entropy, defined in Eqs.~\eqref{fourier_def_2} and \eqref{z_S_omega} 
is computed as
\begin{align}
    S_\omega 
        &= \frac{s}{ 2 i \sqrt{2\pi}} \int_{- \pi n /\kappa }^{ \pi n /\kappa } dt \, ( e^{i\kappa t} - e^{-i\kappa t} ) e^{-i\omega t},\label{harmonic_sin_s_omega_n} \\
        &= i (-1)^{n} \sqrt{ \frac{2}{\pi} } \frac{ s\kappa }{ \kappa^2 - \omega^2 } \sin( \frac{\pi n \omega}{\kappa} ),\notag \\
        &= i \sqrt{ \frac{\pi}{2} } s \left[ \frac{ \sin( \pi n (\omega + \kappa) / \kappa ) }{\pi(\omega + \kappa)} - \frac{ \sin( \pi n (\omega - \kappa) / \kappa ) }{\pi(\omega - \kappa)} \right].\notag
\end{align}
Note that there is no singularity at $\omega = \pm \kappa$, as the vanishing sine cancels the zero in the denominator:
\begin{equation}
    \lim_{\omega \to \pm \kappa } S_\omega = \mp i \sqrt{ \frac{\pi}{2} } \frac{n s}{ \kappa }.
\label{S_omega_lim_harmonic_sin}
\end{equation}

The Fourier transform $S_\omega$ in Eq.~\eqref{harmonic_sin_s_omega_n} falls off at large frequencies as $\sim 1/\omega^2$, and it is smooth at $\omega \to 0$.
This ensures that the total radiated energy and the particle count will be both UV and IR convergent, as follows from Secs.~\ref{sec:IR_div_purity} and \ref{sec:UV_disc}.

%%%%%%%%%%%%%%%%%%%
%%%%%%%%%%%%%%%%%%% 
\begin{figure}[h]
    \centering
    \includegraphics[width=0.9\columnwidth]{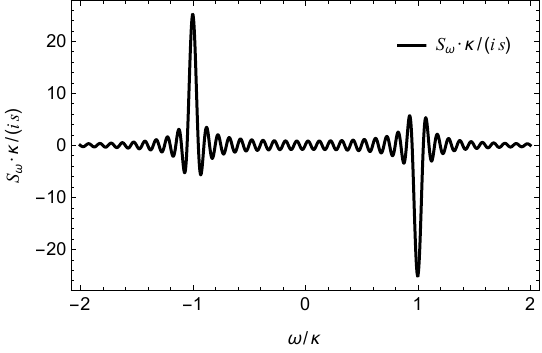}
    \caption{ $S_\omega$ from Eq.~\eqref{harmonic_sin_s_omega_n}, for the finite-oscillations  harmonic mirror. For this plot $n=20$. 
    }
\label{fig:harmonic_SIN_S_omega}
\end{figure}
%%%%%%%%%%%%%%%%%%%
%%%%%%%%%%%%%%%%%%%

\begin{figure}[h]
    \centering
    \includegraphics[width=0.99\columnwidth]{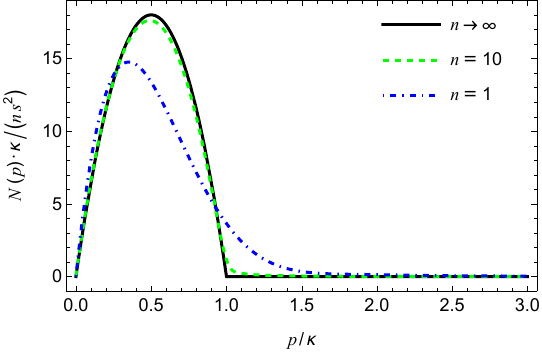}
    \caption{ Normalized particle distribution $N(p) \cdot \kappa/ (n s^2)$ over $n$  oscillations of the finite-oscillations harmonic mirror.
        The solid black line represents the large-$n$ estimate Eq.~\eqref{Np_harmonic_SIN_3}, 
while the dashed and the dot-dashed lines are the numerical results for Eq.~\eqref{particle_distr_from_S} 
    at $n=10$ and $n=1$ respectively.
    }
\label{fig:harmonic_SIN_N_p_1}
\end{figure}

While the resulting particle spectrum $N(p)$ can be evaluated analytically, the resulting expression is long and not particularly revealing. Instead let us consider the limit of large $n$. 
In this case, $S_\omega$ becomes sharply peaked near $\omega = \pm \kappa$ as shown in Fig.~\ref{fig:harmonic_SIN_S_omega}; 
in the limit $n\to\infty$ it becomes a sum of two delta functions, 
\begin{equation}
    S_\omega \xrightarrow[n\to\infty]{} \sqrt{ \frac{\pi}{2} } i s \left[  \delta(\omega + \kappa) - \delta(\omega - \kappa) \right].
\label{S_delta_harmonic_SIN}
\end{equation}
Then to leading order in large $n$, Eqs.~(\ref{particle_distr_from_S}) and (\ref{harmonic_sin_s_omega_n}) yield the particle number distribution
\begin{equation}
    N(p) \approx \begin{cases}
        \dfrac{72 s^2 p (\kappa - p) }{ \kappa^3 }\ n  \,, \quad &0 < p < \kappa, \\
        0 \,, \quad &p > \kappa.
    \end{cases}
\label{Np_harmonic_SIN_3}
\end{equation}
This large $n$ limit turns out to work quite reasonably for $n=10$ and not too poorly for $n=1$, as seen in Fig.~\eqref{fig:harmonic_SIN_N_p_1}.

%%%%%%%%%%%%%%%%%%%
%%%%%%%%%%%%%%%%%%% 
%%%%%%%%%%%%%%%%%%%
%%%%%%%%%%%%%%%%%%%

\begin{figure}[h]
    \centering
    \includegraphics[width=0.9\columnwidth]{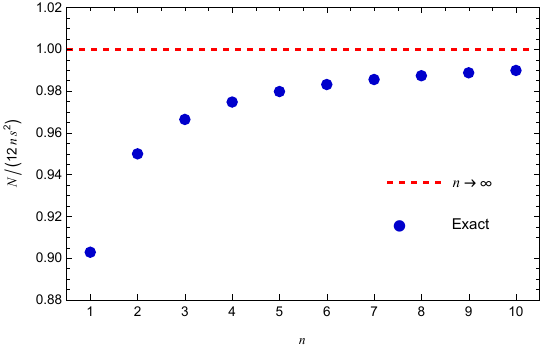}
    \caption{ Exact (numerical) ratio $N  / 12ns^2$ for different integers $n$, for the finite-oscillations  
    harmonic mirror. The red dashed line corresponds to the large-$n$ limit  Eq.~\eqref{harmonic_SIN_energy_particles}, while the blue dots are the result of numerical integration in Eq.~\eqref{particle_distr_from_S}.
    }
\label{fig:harmonic_SIN_Ntot_ratio_exact}
\end{figure}

With the explicit form of $N(p)$ at hand, we can easily compute the total particle count and the total radiated particle-energy: 
\begin{equation}
\begin{aligned}
    N  &= \int_0^\infty \diff p \, N(p) \approx 12ns^2, \\
    E  &= \int_0^\infty  \diff p \, p \,  N(p) = 6 n s^2 \kappa.
\end{aligned} \label{harmonic_SIN_energy_particles}
\end{equation}
From Fig.~\ref{fig:harmonic_SIN_Ntot_ratio_exact} one can see that the exact total particle count $N $ indeed approaches the large-$n$ asymptotics \eqref{harmonic_SIN_energy_particles}.
As for the total energy, it can be checked with Eq.~\eqref{stress_energy}, which gives $E  = 6 n s^2 \kappa$ exactly for any integer $n$, as no singular contributions are coming from the boundary.
We have checked that indeed  $E  = 6 n s^2\kappa$ exactly for all $n$ within the numerical machine error.

One can see that the quantities Eqs.~\eqref{harmonic_SIN_energy_particles} are extensive in $n$, as expected.
What is interesting, however, is that even at large $n$, i.e., after a large number of oscillations, the radiated spectrum Eq.~\eqref{Np_harmonic_SIN_3} is centered around $\kappa/2$ in the frequency space [and not $\kappa$, which is the oscillator's frequency, Eq.~\eqref{z_t_harmonic_sin}], and retains a fixed non-vanishing width. 
This observation is directly related to the following property: at large $n$, % that $E /N =\hbar\kappa/2$.  
\begin{equation}
    E  \approx \frac{ \hbar \kappa}{2} N  \,,
\label{E_N_relation_sin}
\end{equation}
where we have reinstated $\hbar$ for illustrative purposes.
Eq.~\eqref{E_N_relation_sin} shows that the average energy per quantum is 
$\hbar \kappa / 2$, consistent with the fact that the emission spectrum 
is peaked around frequency $\kappa/2$.

%%%%%%%%%%%%%%%%%%%
%%%%%%%%%%%%%%%%%%% 
%%%%%%%%%%%%%%%%%%%
%%%%%%%%%%%%%%%%%%%

%%%%%%%%%%%%%%%%%%% 
\subsection{Damped oscillator} 
\label{sec:harmonic-damp} 

Another interesting time-dependent entropy corresponds to the damped harmonic oscillator:
\begin{equation}
S(t) = \begin{cases} 
    s e^{- \frac{ \kappa t }{ 4 \pi n} } \sin(\kappa t) , \quad &t > 0, \\
    0 \,, \quad &t <0.
    \end{cases}
\label{S_t_harmonic_damped}
\end{equation}
Despite allowing $t\to\infty$ for the oscillator, the exponential factor effectively cuts the number of full periods of motion at the order $O(n)$.
The coefficient in the exponent is chosen so as to relate to the results in Eq.~\eqref{harmonic_SIN_energy_particles}, see below.

\begin{figure}[h]
    \centering
    \includegraphics[width=0.9\columnwidth]{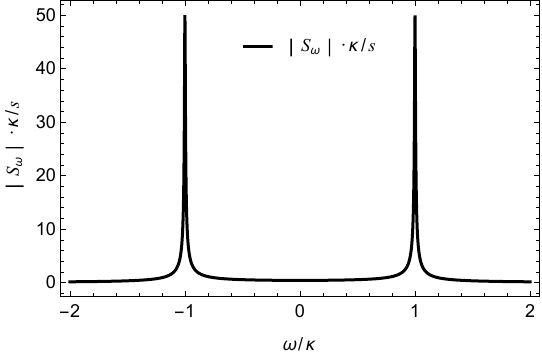}
    \caption{ Absolute value of $S_\omega$ from Eq.~\eqref{S_omega_damped}, for the damped harmonic mirror. For this plot $n=20$. 
    }
\label{fig:harmonic_DAMP_S_omega}
\end{figure}

In this case, the Fourier transform is given by
\begin{equation}
    S_\omega = \frac{s \kappa}{ \sqrt{2\pi} } \frac{1}{ (\kappa^2 - \omega^2) + i \omega \frac{\kappa}{2 \pi n} + \frac{ \kappa^2 }{ 16 n^2 \pi^2 } }\ .
\label{S_omega_damped}
\end{equation}
Note that the total particle and energy integrals will also be UV and IR finite.  Figure~\ref{fig:harmonic_DAMP_S_omega} plots this entropy.

%%%%%%%%%%%%%%%%%%%
%%%%%%%%%%%%%%%%%%% 
%%%%%%%%%%%%%%%%%%%
%%%%%%%%%%%%%%%%%%%

At large $n$, the last term in the denominator of Eq.~\eqref{S_omega_damped} can be dropped, while the imaginary term yields a pole regularization prescription\footnote{This regularization is different from the usual $i\epsilon$ prescription in QFT, as in the present case, both poles in $\omega$ are above the real axis.}.
Thus by the Sokhotski–Plemelj theorem we obtain
\begin{equation}
    S_\omega \xrightarrow[n\to\infty]{} \frac{s \kappa}{ \sqrt{2\pi} } \left( - i \pi \delta(\kappa^2 - \omega^2) + \mathcal{P} \frac{1}{\kappa^2 - \omega^2} \right)\ ,
\end{equation}
where $\mathcal{P}$ is the Cauchy principal value.

\begin{figure}[h]
    \centering
    \includegraphics[width=0.99\columnwidth]{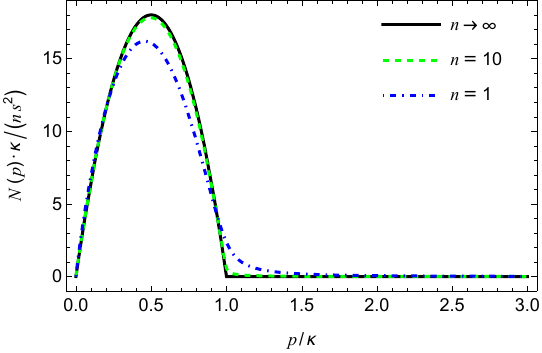}
    \caption{ Ratio $N(p) / n s^2$ over $n$ oscillations of the damped harmonic mirror.
        The solid black line represents the large-$n$ estimate Eq.~\eqref{Np_harmonic_DAMP_3}, while the dashed and the dot-dashed lines are the numerical results for Eq.~\eqref{particle_distr_from_S} at $n=10$ and $n=1$ respectively. 
    }
\label{fig:harmonic_DAMP_N_p_1}
\end{figure}

Now, we will apply the same logic as before to estimate the particle integral, substituting the $n \to \infty$ limit for one of the $S_\omega$ in $|S_\omega|^2$.
The $\mathcal{P}$ integral turns out to vanish, while the $\delta$-function integral produces
\begin{equation}
    N(p) \approx \begin{cases}
        \dfrac{72 s^2 p (\kappa - p) }{ \kappa^3 }\ n  \,, \quad &0 < p < \kappa, \\
        0 \,, \quad &p > \kappa \,,
    \end{cases}
\label{Np_harmonic_DAMP_3}
\end{equation}
which is the same as in Eq.~\eqref{Np_harmonic_SIN_3} from the previous model.
Figure~\ref{fig:harmonic_DAMP_N_p_1} plots this particle spectrum.
From this we straightforwardly obtain
\begin{equation}
\begin{aligned}
    N  &= \int_0^\infty \diff p \, N(p) \approx 12 ns^2, \\
    E  &= \int_0^\infty \diff p \, p \,  N(p) = 6 n s^2 \kappa.
\end{aligned} 
\label{harmonic_DAMP_energy_particles}
\end{equation}
which is again the same as the result Eq.~\eqref{harmonic_SIN_energy_particles} for the finite-oscillations mirror.
The result for $N $ is close to the numerical result at large $n$.
At the same time, the formula for the total radiated energy can be compared with Eq.~\eqref{stress_energy}, which gives $E  = 3 n s^2 \kappa$ exactly.
We see that the approximate formula $ E  \approx \frac{1}{2} \hbar \kappa N  $ still holds (reinstating $\hbar$).

%%%%%%%%%%%%%%%%%%%
%%%%%%%%%%%%%%%%%%% 
%%%%%%%%%%%%%%%%%%%
%%%%%%%%%%%%%%%%%%%

%%%%%%%%%%%%%%%%%%%%%%% 
\subsection{Note on  discontinuous entropy and UV divergencies}

Briefly, consider another interesting case: shift the phase of the oscillator in Eq.~\eqref{z_t_harmonic_sin} by a quarter of a period, while keeping the same time frame.
This seemingly harmless operation makes the velocity and the entropy discontinuous; instead of Eq.~\eqref{S_t_harmonic_sin} we are going to have
\begin{equation}
S(t) = \begin{cases} 
    s \cos(\kappa t) , \quad &t \in [-\pi n /\kappa, \pi n /\kappa], \\
    0 \,, \quad &t \notin [-\pi n /\kappa, \pi n /\kappa].
    \end{cases}
\label{S_t_harmonic_DISCONT}
\end{equation}
Its Fourier transform turns out to decay only as $\sim 1 / \omega$ at large frequencies, which makes the total particles integral logarithmically divergent in the UV, while the total energy integral is going to be linearly divergent.
From the point of view of the stress-energy integral Eq.~\eqref{stress_energy}, the energy is not well-defined due to the boundary contributions at $t = \pm \pi n /\kappa$, where $S_t'$ contains $\delta$-function terms and $(S_t')^2$ is not a well-defined quantity.

What is interesting, however, is that, point-wise, the particle number distribution $N(p)$ still converges to Eq.~\eqref{Np_harmonic_SIN_3} at large $n$.
The $1/p$ tail of $N(p)$ turns out to be subleading at $n \to \infty$.

Taking the limit of large number of oscillations in the discontinuous case Eq.~\eqref{S_t_harmonic_DISCONT} effectively renders the boundary effects to be subleading; therefore, the related UV divergencies are not seen in this limit, and we recover the same results as for the UV-finite models, see e.g.\ Eqs.~\eqref{harmonic_SIN_energy_particles} and \eqref{harmonic_DAMP_energy_particles}.

This illustrates the point of Sec.~\ref{sec:UV_disc}.
The UV divergencies here are related only to microscopical details of the trajectory or the entropy.
They can be regularized.
The long-wavelength physics, independent of the microscopic details, is the same for different regularization schemes, either in hard-cutoff regularization of Sec.~\ref{sec:harmonic-fin}, or in exponential regularization of Sec.~\ref{sec:harmonic-damp}.

\section{Conclusions} \label{sec:concl} 

We presented a framework for understanding the interplay between entanglement entropy and particle creation. 
Entanglement, quantified by  $S(t)$, is a key ingredient, at least in the moving mirror model, even while the mirror’s motion is currently understood as the primary driver of particle production. We have demonstrated this for accelerating mirrors, black hole evaporation, and beta decay. The dependence of the particle spectrum $N(\omega)$ on \( S(t) \) suggests that, in principle, manipulating entanglement entropy could lead to particle creation when a mechanism exists to change \( S(t) \).

This raises the intriguing possibility of particle production in systems without acceleration, driven purely by time-dependent entanglement structure. Conceptually similar mechanisms have been proposed in media with time-dependent dispersion and dissipation, where radiation can be generated without spatial motion of boundaries~\cite{Lang:2019avs}.

The conceptual premise for this link echoes Wheeler’s notion that “it from bit,” suggesting informational dynamics alone may suffice to generate quanta. We explored this mechanism in novel contexts, including the non-relativistic decay of the neutron and a quantum pure finite-particle black hole analog. In addition, we extended it to scenarios of large particle production induced by harmonic time dependence and addressed the conditions required to have well-behaved IR and UV limits.

Future investigations could explore potential advantages of the entropy-based formulation over the motion-based approach in dimensions higher than 1+1. Another promising direction is to examine the entanglement–creation link in regimes involving not only large particle production but also substantial instantaneous entropy, the analog of relativistic motion.

\begin{acknowledgments}
M.G. is supported, in part, by the FY2024-SGP-1-STMM Faculty Development Competitive Research Grant (FDCRGP) no.201223FD8824 and SSH20224004 at Nazarbayev University in Qazaqstan. The work of E.I. is supported, in part, by U.S. Department of Energy Grant No. de-sc0011842.
\end{acknowledgments}

\appendix

\section{Step function Fourier transform}
\label{sec:math}

Here we demonstrate the validity of Eq.~\eqref{S_omega_IR_div}. For a step function
\begin{equation}
    f(t) = \begin{cases}
        A \,, \quad &t < 0, \\
        B \,, \quad &t > 0,
    \end{cases}
\label{step_func_t}
\end{equation}
we can compute its Fourier transform in the $\varepsilon$-regularization standard in QFT.
Take some $\varepsilon > 0$.
Consider the two following integrals:
\begin{equation}
\begin{aligned}
    I_+ &= \int_0^\infty \diff{t} e^{- i \omega t - \varepsilon t} = \frac{1}{i \omega + \varepsilon}, \\
    I_- &= \int_{-\infty}^0 \diff{t} e^{- i \omega t + \varepsilon t} = \frac{1}{ - i \omega + \varepsilon}. \\
\end{aligned}
\end{equation}
We have
\begin{equation}
\begin{aligned}
    I_+ - I_- &= \frac{ - 2 i \omega }{\varepsilon^2 + \omega^2 } \xrightarrow[ \varepsilon \to 0 ]{ } - \frac{ 2 i }{ \omega }, \\
    I_+ + I_- &= \frac{2 \varepsilon }{\varepsilon^2 + \omega^2 } \xrightarrow[ \varepsilon \to 0 ]{ } 2 \pi \delta(\omega).
\end{aligned}
\end{equation}
The first of these limits is trivial, while the second is the basic property of the Poisson kernel.

Using these, we can write for the Fourier transform of $f$:
\begin{align}
    \sqrt{2\pi} f_\omega 
        &= A I_- + B I_+, \notag\\
        &= \frac{1}{2} (A + B) (I_+ + I_-) - \frac{1}{2} (A - B) (I_+ - I_-), \notag \\
        &= (A + B) \pi \delta (\omega) + (A-B) \frac{ 2 i }{ \omega }.
\end{align}
In the context of this paper, the delta-function term can be dropped.
$A-B$ is the difference of $f(t)$ on $t \to \pm \infty$.

The same result holds in the IR limit $\omega \to 0$ for an arbitrary function that asymptotes to Eq.~\eqref{step_func_t} at $t \to \pm \infty$, up to an insignificant constant phase factor.
This proves Eq.~\eqref{S_omega_IR_div}.

\section{Example of analytic particle production from a smooth, slow mirror}\label{sec:slow_smooth_mirror}

This appendix presents an example, similar in form to the previous one in section \ref{sec:lorentzian_mirror}, to help further illustrate how the entropic approach can provide an analytic solution for finite particle emission from a perfectly reflecting mirror. We consider the time-dependent trajectory
$z(t) = -\kappa^{-1}\,\tan^{-1}(e^{\kappa t})$
originally studied in a one-sided form in~\cite{good2013time}, where it was used to investigate finite particle production.  That analysis was limited to a single side of the mirror and the speed was fixed at half the speed of light.  Using both sides of the mirror, and multiplying $z$ by a constant $2|v_\textrm{max}|$,
\begin{equation}
    z(t) = -2\frac{|v_\textrm{max}|}{\kappa}\,\tan^{-1}(e^{\kappa t}),
\end{equation}
guaranteeing strictly small velocities and small entanglement, the power $\alpha(t)^2/6\pi$ leads to the total energy:
\be E = \frac{\kappa}{24\pi}\left[\frac{\gamma ^3 \sin ^{-1}v}{v}+ \gamma ^2-2\right] \approx \frac{\kappa v^2}{9 \pi }+\mathcal{O}\left(v^4\right), \label{energy_artx}\ee
where we have dropped the subscript on $v_\textrm{max}$ and used the magnitude of the maximum velocity and its Lorentz factor $\gamma^{-2} = 1-v^2$. 

The motivation here is to leverage the physical properties of the Arctx mirror above and simplify the analytical results for particle production.

The trajectory is asymptotically inertial,
\begin{equation}
    \lim_{t \to \pm\infty} \dot{z}(t) = 0,
\end{equation}
ensuring finite energy and particle production. The magnitude of the von Neumann entanglement entropy will then be
\begin{equation}
    S(t) = -\frac{\dot{z}(t)}{6}  = \frac{v}{3} \ \frac{e^{\kappa t}}{1 + e^{2\kappa t}} = \frac{v}{6} \, \text{sech}(\kappa t),
\end{equation} 
which peaks at \( S_{\max} = v/6 < 1/6 \), satisfying the causality constraint for entanglement-driven radiation with a suitable $v$ much smaller than 1. 

The particle production is obtained from the Fourier transform, $S_\omega = \mathcal{F}_\omega S(t)$, 
\begin{equation}
    |S_\omega|^2 = \frac{\pi v^2}{72\kappa^2} \  \text{sech}^2\left(\frac{\pi  \omega }{2\kappa}\right).
\end{equation}

The total particle count, Eq.~(\ref{totalparticles}), is given by
\begin{equation}
    N = \frac{144}{\pi} \int_0^\infty \int_0^\infty \frac{p q}{(p + q)^2} |S_\omega|^2 \diff{p} \diff{q} = \frac{4v^2}{3 \pi^2} \ln 2, \label{N_arctx}
\end{equation}
Similarly, the total radiated energy, Eq.~(\ref{particle_energy}), is
\begin{equation}
    E = \frac{144}{\pi} \int_0^\infty \int_0^\infty \frac{p^2 q}{(p + q)^2} |S_\omega|^2 \diff{p} \diff{q} = \frac{\kappa v^2}{9\pi}. \label{E_arctx}
\end{equation}
This result agrees with the stress-energy, $E=\kappa v^2/(9\pi)$, of Eq.~(\ref{energy_artx}) and calculated directly from a derivative of the entropy, Eq.~(\ref{stress_energy}).
The entropic construction provides a clean analytical solution for finite particle and energy emission.

\section{Temperature by Entanglement}

It is natural to consider whether one can connect the entropy that is the main focus of this article to a temperature. 
The first law of thermodynamics gives 
\begin{equation}
\diff{Q} = T\, \diff{S}, \label{eq:first_law}
\end{equation}
where \( \diff{Q} \) is the infinitesimal heat lost, \( T \) is the temperature, and \( \diff{S} \) is the change in entropy.

The entropy part, for a non-relativistic system where we here restore SI units, associated with motion is
\begin{equation}
    S = -\frac{\dot{z}}{6} \frac{k_B}{c} \quad \Rightarrow \quad \diff{S} = -\frac{\ddot{z}}{6} \frac{k_B}{c} \, \diff{t},  \label{eq:entropy}
\end{equation}
where \( \dot{z} \) and \( \ddot{z} \) denote the velocity and acceleration, respectively.

The heat part can be related to the total radiated power via 
\begin{equation}
    P = \frac{\hbar\, \ddot{z}^2}{6\pi c^2}\ ,  \label{eq:power}
\end{equation}
so that the energy (heat) lost to one side in time \( \diff{t} \) is
\begin{equation}
    \diff{Q} = -\frac{P}{2}\diff{t} = -\frac{\hbar\, \ddot{z}^2}{12\pi c^2} \diff{t}.  \label{eq:heat_loss}
\end{equation}
Combining Eq.~\eqref{eq:entropy} and Eq.~\eqref{eq:heat_loss} in the first law Eq.~\eqref{eq:first_law} gives the effective temperature:
\begin{equation}
    T = \frac{\diff{Q}}{\diff{S}} = \frac{\hbar\, \ddot{z}}{2\pi c\, k_B}.  \label{eq:temperature}
\end{equation}
This result suggests an acceleration thermality interpretation associated with acceleration-induced radiation. This complements the acceleration thermality previously outlined via beta decay in section \ref{sec:beta_decay}, cf.~Eq.~(\ref{thermal_acceleration_particle_spectrum}).

\section{Large displacement validity}
\label{sec:expansion_beta_check}

Here we present an example with an infinite-displacement trajectory, demonstrating the non-relativistic expansion formula for the beta Bogolyubov coefficients (see Eq.~(85) of \cite{Mujtaba:2024vmf})
\begin{equation}
	\beta_{p q}^R = \sqrt{ \frac{2pq}{\pi} }\  \sum_{n=1}^\infty \frac{ i^n \omega_-^{n-1} }{n!} \cdot \mathcal{F}\left[ z(t)^n \right](\omega),
\label{beta_all_order}
\end{equation}
remains valid for large displacements. We take the trajectory $z(t)$ as in Eq.~\eqref{betadecay_traj}. In this appendix only, we take all frequencies in units of $\kappa$ to simplify comparison with the cited works. 

On one hand, since the beta Bogolyubov coefficients in this case are known exactly (see e.g. Eq.~(11) of \cite{Good:2016yht}\footnote{Note that $\beta_{p q}^R$ in \cite{Good:2016yht} has an extra overall minus sign due to a slightly different contour convention; however, this is not important, as only $|\beta_{p q}^R|^2$ is physically significant. Also, in our convention here the mirror goes to $z\to +\infty$ at late times.}), we can directly expand them in a power series in the speed $s$.
We start from the expression
\begin{equation}
	\beta_{pq}^R = - \frac{ e^{- \omega \pi / 2 } s }{\pi }  \sqrt{pq} \cdot ( \omega - s \omega_- )^{i \omega - 1} \cdot \Gamma (- i \omega) \,,
\label{betaR_exact_2}
\end{equation}
where $\omega = p + q$ and $\omega_- = p - q$.
With the help of the generalized binomial theorem and repeatedly using the identity $x \Gamma(x) = \Gamma (x+1)$, we can expand Eq.~\eqref{betaR_exact_2} as
\begin{equation}
	\beta_{pq}^R
		= - \frac{ \sqrt{pq} e^{- \omega \pi / 2 } }{\pi} \ \sum_{n=0}^\infty \frac{ i s^{n+1} \omega_-^n  }{   n! }  \omega^{i \omega - n - 2}\  \Gamma (- i \omega + n + 1) \,.
\label{beta_all_order_check_1}
\end{equation}

On the other hand, we can apply the expansion formula \eqref{beta_all_order} with the trajectory from Eq.~\eqref{betadecay_traj}.
It involves computing the Fourier integrals
\begin{equation}
	\mathcal{F}\left[ z(t)^n \right](\omega)
		= \frac{1}{\sqrt{2\pi}} \int\limits_{-\infty}^{\infty} \diff t \, (s W(e^t))^n \, e^{-i \omega t } \,.
\end{equation} 
These integrals can be computed by passing from the integration variable $t$ to $W \equiv W(e^t)$ according to
\begin{equation}
	t = W + \ln W \,, \quad
	dt = (1 + 1/W) dW
\end{equation}
and using Appendix~C of \cite{Ievlev:2023inj}.
Simplifying the result with $x \Gamma(x) = \Gamma (x+1)$, we obtain
\begin{equation}
	\mathcal{F}\left[ z(t)^n \right](\omega) = n \frac{ -i ( - i s)^n }{ \sqrt{2\pi} }  e^{- \omega \pi / 2 }
				\omega^{ i \omega - n - 1 }  \, \Gamma (- i \omega + n  ) \,.
\label{beta_z_fourier_3}
\end{equation}
Finally, plugging Eq.~\eqref{beta_z_fourier_3} into Eq.~\eqref{beta_all_order}, we arrive at precisely the same series as in Eq.~\eqref{beta_all_order_check_1}.

This example demonstrates that the expansion formula appears to be effective even for asymptotically inertial trajectories, see e.g., \cite{Fabbri:2005mw}, beyond the constraint of asymptotic rest, Eq.~(\ref{smooth}), or a turned-over Page curve, Eq.~(\ref{smoothS}). That is, the formula works even for systems with infinite soft particle production, e.g., Eq.~(\ref{soft}).

\bibliography{main}

\end{document}